\documentclass[a4paper,20pt]{article}
    \usepackage[T1]{fontenc}
    \usepackage{graphicx}
    \usepackage{amsmath}
    \usepackage{amsfonts}
    \renewcommand{\abstract}{}
\begin{document}
\makeatletter
\renewcommand{\@oddhead}{\textit{YSC'14 Proceedings of Contributed Papers} \hfil \textit{K. Bryndal, B. Wszo\l ek}}
\renewcommand{\@evenfoot}{\hfil \thepage \hfil}
\renewcommand{\@oddfoot}{\hfil \thepage \hfil}
\fontsize{11}{11} \selectfont

\title{Spectroscopic Families Among Diffuse Interstellar Bands}
\author{\textit{K. Bryndal, B. Wszo\l ek}}
\date{}
\maketitle
\begin{center} {Jan D\l ugosz Academy, Institute of Physics, al. Armii
Krajowej 13/15, 42-200 Cz\k{e}stochowa, Poland \\ bogdan@ajd.czest.pl}
\end{center}

\begin{abstract}
\indent Looking for spectroscopic families in the whole set of discovered
diffuse interstellar bands (DIBs) is an indirect trial of solving
the problem of DIBs' carriers. Basing on optical high resolution
spectra, covering the range from 5655 to 7020 \AA, we found few
relatively strong DIBs which are not well correlated one with
another and therefore they may play a role of representatives of
separate spectroscopic families. In the next step we indicated
DIBs which tend to follow the behaviour of their representatives.
As a result of our analysis we propose few, probably not complete yet,
 spectroscopic families of DIBs.
\end{abstract}

\section*{Introduction}

\indent \indent DIBs are intriguing absorption structures of interstellar origin,
observed in the spectra of reddened stars of early spectral types.
 The first two bands of this kind were described by Heger in 1922.
 DIBs, as we know them today, are scattered within the whole region
of visible light and in the near infrared. For the extended review
of the topic see e.g. \cite{herbig}. Many DIB carriers have been
proposed over the years. They ranged from dust grains to free
molecules  of very different sizes and structures
\cite{herbig_leka}. Individual DIBs differ strongly between
themselves in intensity, line width and profile shapes.

\indent There is a respectable body of opinion that a single carrier
can not be responsible for all known DIBs \cite{krelovski88}.
 The evidence that DIBs do not originate in a single carrier raised
the question of whether we can divide them into sets characterized
by constant strength ratios. Such sets might then be the spectral
signature of a single carrier which would facilitate an identification
 of the DIB carriers. Kre\l owski and Walker \cite{krelovski87} attempted to group
 the observed DIBs into `families', with each set of features having
constant strength ratios and thus having a common origin.  However,
the proposed families have not been clearly confirmed by the recent
observations. The correlations among the strong DIBs usually suffer
some scatter, which suggests that no pair of these strong DIBs is of
 the same origin.

\indent The general question of DIB families remains an important one, but
with ambiguous results so far; it seems clear that DIBs within certain
 groups correlate better internally than with nonmembers of the group,
but at the same time the correlations are not perfect as might be
expected for features arising from a truly common carrier. The
recent discoveries of numerous very weak DIBs \cite{galazutdinov}
provides a new challenge and new opportunity to seek bands that have
common carriers.

\indent Advanced study of the problem of spectroscopic families was
done by Wszo\l ek and God\l owski \cite{wszolek}. The authors
analysed relatively strong DIBs as well as very weak ones. They
applied three different methods to separate spectroscopic families
of DIBs and explored the best approach for future investigations of
this type. On the basis of their methods authors suggested some
candidates as spectroscopic `relatives' of 5780 and 5797 \AA. Weak
DIBs  5776 and 5795  tend to join 5780 \AA \,band, and 5793, 5818,
5829, 5850 bands seem to belong to spectroscopic family of 5797 \AA.
Using the same methods as Wszo\l ek and God\l owski \cite{wszolek}
we continue our search for spectroscopic families and here we
present some of more interesting results.

\section*{Observational Data}
\indent \indent All the spectra analysed for the purpose of this
contribution were taken from the archives of  Kre\l owski
(Astronomical Centre, Nicolaus Copernicus University in Toru\'n,
Poland). We used spectra acquired at the McDonald Observatory with
an echelle spectrograph fed with the 2.1-m telescope, covering the
spectral range from 5655 to 7020 \AA \ \cite{krelovski93}. Spectra
have S/N ratios of approximately  200 and a resolution R of about 64
000. We have used spectra for 74 target stars.

\section*{Data Analysis}
\indent \indent After the initial routine spectra reduction, like continuum division or
 dividing by standard to remove telluric lines, we have measured EWs and
depths for several DIBs covering spectral range from 5600 to 7000 \AA.
 Then we drew correlation diagrams for pairs of bands and we calculated
correlation coefficients. There exist many arguments that DIBs 5797
and 5850 have the same carrier. Almost all procedures isolating
DIBs' families indicate that 5797 and 5850 bands tend to belong to
the same family. Figures 1-2 are to show how good is correlation of
5850 with 5797 and how bad it is with 5780 \AA \,band.

\indent We have done dozens of similar correlation diagrams and
looked for pairs of DIBs which are relatively well mutually
correlated. We found that strong DIBs: 6196, 6203, 6270 and 6614 are
very good mutually correlated (coefficient of correlation for pairs
of them usually equals 0.99). They all are quite well correlated
also with 5780 band. We have found that strong DIB at  6379 \AA \,is
relatively bad correlated with 5780 and 5797 DIBs. It is also bad
correlated with 6196 and 6203 DIBs. We have also learnt that one has
to be very careful using only formal correlation between considered
DIBs as a criteria of belonging to any spectroscopic family. When we
correlate EWs of strong and very weak DIBs,  correlation may be much
worse due to measuring errors. Probably this is why intensity of
strong 5797 DIB is not strictly correlated with its much
 weaker `relative' - 5850 band (see Fig. 2). These two DIBs most probably
belong to the same spectroscopic family and they are not strictly
correlated.

\section*{Discussion}
\indent \indent To isolate spectroscopic families one needs very
good observational data and one has to do very precise spectral
analysis. As a result of our investigations, we qualified 5780,
6196, 6203, 6270 and 6614 DIBs as good candidates to the same
spectroscopic family. The 5780 DIB is the weakest candidate here. If
5780 does not belong to this family, it has to be a representative
of separate spectroscopic family.

\indent Studying correlation diagrams we have also found that a strong DIB at
6379 \AA \, is a very probable representative of the new spectroscopic family.
 More extensive investigations, being currently in progress, will allow
us to indicate probable family members joining their representative.

\section*{Acknowledgements} We would like to thank Professor Kre\l owski
for giving us an access to his data archives.

\newpage

\textbf{Figure 1.} Correlation diagram for equivalent widths of 5850
and 5780 DIBs. \vspace{10ex}

\textbf{Figure 2.} Correlation diagram for equivalent widths of 5850
and 5797 DIBs. \vspace{10ex}

Figures are available on YSC home page
(http://ysc.kiev.ua/abs/proc14$\_$4.pdf).

\end{document}